\documentclass[times]{article}
\usepackage{url}
\usepackage{amsmath,amsfonts,amssymb,bm}
\usepackage{hyperref}
\usepackage{graphicx}
\usepackage{booktabs}
\usepackage{multirow}
\usepackage{authblk}
\usepackage{geometry}

\title{Distilled Non-Semantic Speech Embeddings with Binary Neural Networks for Low-Resource Devices}

\author[1]{Harlin Lee\thanks{harlin@math.ucla.edu. Funding: Harlin Lee's work was partially supported by NSF DMS-1952339.}}

\author[2]{Aaqib Saeed\footnote{a.saeed@tue.nl.}} 

\affil[1]{Department of Mathematics, University of California Los Angeles}
\affil[2]{Eindhoven University of Technology}

\begin{document}

\maketitle

\begin{abstract}
This work introduces BRILLsson, a novel binary neural network-based representation learning model for a broad range of non-semantic speech tasks. We train the model with knowledge distillation from a large and real-valued TRILLsson model with only a fraction of the dataset used to train TRILLsson. The resulting BRILLsson models are only $2$MB in size with a latency less than $8$ms, making them suitable for deployment in low-resource devices such as wearables. We evaluate BRILLsson on eight benchmark tasks (including but not limited to spoken language identification, emotion recognition, human vocal sounds, and keyword spotting), and demonstrate that our proposed ultra-light and low-latency models perform as well as large-scale models.\\

\end{abstract}

\section{Introduction}

Representation learning takes advantage of large amounts of unlabeled data to learn features that can be used for a variety of downstream signal processing and machine learning tasks. This has demonstrated especially impressive performance in speech and audio processing  \cite{saeed2021contrastive, shor20_interspeech, shor2022trillsson}, as the ubiquity of smart phones, watches, and home appliances has made it easy and inexpensive to collect a wealth of unlabeled audio signals. In particular, there is growing interest in using representation learning to build general-purpose models for non-semantic speech tasks, which are problems related to human speech other than its meaning, such as spoken language identification \cite{maclean2018voxforge}, emotion recognition \cite{cao2014crema,SCHONEVELD20211}, human vocal sounds \cite{gong_vocalsound}, keyword spotting \cite{warden2018speech} and more. 

However, the large sizes of the trained models and the amount of computational resources that are required to run them on newly acquired data have stalled the real-world deployment of existing models for non-semantic speech applications. These assumptions are critical in mobile computing, edge computing, internet-of-things (IoT), and tiny machine learning (tinyML) settings, which are where many paralinguistic speech applications actually take place, e.g., in small wearable for healthcare or with voice-controlled artificial intelligence (AI) assistants in smart devices. Although models such as FRILL \cite{peplinski21_interspeech} and TRILLsson \cite{shor2022trillsson} were recently proposed to reduce the complexity and size of the deep learning models, a large gap still remains between \textit{highly effective} (i.e., accurate) and \textit{highly efficient} (i.e., light enough to be run on devices as small as smartwatches) representation learning models for paralinguistic speech tasks.

To this end, we design and evaluate BRILLsson, binary neural networks (BNNs) \cite{courbariaux2015binaryconnect} that are small and fast enough to be deployed in devices with limited memory and computational resources. BNNs have weights of only +1 or -1, which make them ideal compact architectures, especially in conjunction with co-designed machine learning hardware that one may see in modern IoT applications. Furthermore, we employ knowledge distillation~\cite{hinton2015distilling,Romero15-iclr, ZARAS2021215} from TRILLsson to BRILLsson, and show that distillation can be achieved using data that is slightly unrelated or smaller than the one used to train the original model, which is beneficial when original data is not available or so large that it requires extensive computing power. Finally, we illustrate that BRILLsson achieves performance on many non-semantic speech benchmark and other tasks that is comparable to that of much larger models.

In summary, our main contributions are:
\begin{itemize}
    \item We propose BRILLsson, ultra-light and fast models for representation learning that are suitable for low-resource devices. BRILLsson's size is only $2$MB, and its latency is less than 8ms.
    \item Perform successful knowledge transfer via embedding distillation from a large-scale real-valued model to binary neural networks. While similar approaches have been explored in image classification \cite{leroux2020training} and 
    speech separation \cite{chen2018distilled}, ours is the first in the context of general-purpose representation learning for non-semantic speech.
    \item Demonstrate that despite their compressed size, our BNNs perform comparably to TRILL, FRILL, and TRILLsson on eight different benchmark datasets.
    \item Our models are ideal for continuous on-device inference for privacy-preserving health monitoring (e.g., coughing, sneezing) due to its low-computational footprint.  
\end{itemize}

\section{Methods}

\begin{figure}[ht]
    \centering
    \includegraphics[width=0.4\columnwidth]{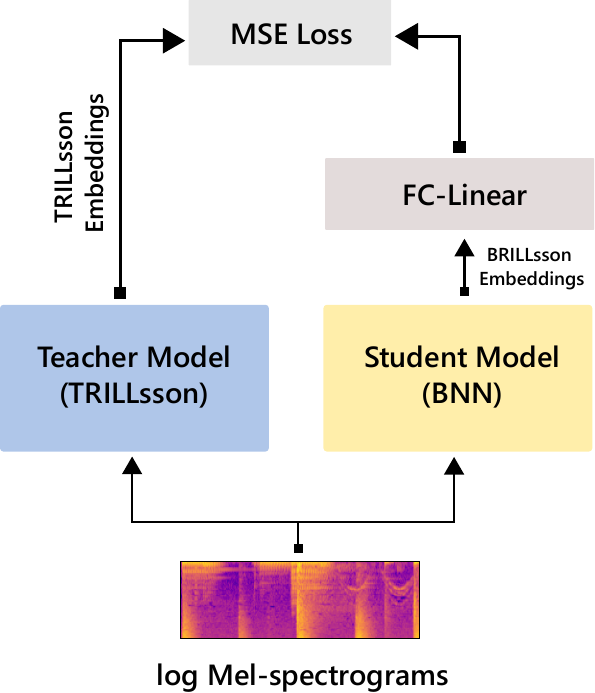}
    \caption{Illustration of distilling binary neural models for non-semantic speech.}
    \label{fig:overview}
\end{figure}

Our objective is to create extremely compact general-purpose audio models that 1) generate informative embeddings (or representations) for a broad range of audio recognition tasks, and 2) can run efficiently on-device with low latency for low-resource devices, e.g., wearables without constant connection to the cloud. 
We use knowledge distillation to transfer learned representations of a large-scale pre-trained teacher model to smaller BNN-based student models that are otherwise difficult to pre-train with self-supervised learning (or a similar strategy) due to their limited capacity. 
We would like to highlight that while distillation has been successfully leveraged for knowledge transfer for non-semantic speech before, to the best of our knowledge, this work is the first attempt at utilizing it for ultra-compact binary neural networks. Figure \ref{fig:overview} provides high-level illustration of our approach, and the following subsections describe its essential building blocks.  

\subsection{Embedding Distillation}
Distillation is a technique to create a model with smaller size and less computational load without sacrificing its effectiveness~\cite{hinton2015distilling}. It transfers information from a large supervising model ---the teacher ($\mathcal{F}_t$)--- to a relatively small model ---the student ($\mathcal{F}_s$)--- with the goal of compression for efficient inference. The teacher is generally a fixed pre-trained network learned with a massive amount of high-quality data, in other words a privileged model. In contrast, the student is a low capacity network that is guided to imitate the output of the teacher. The information-rich signal from the  teacher enables a compact student network to learn important aspects of the data that would otherwise be missed when solely minimizing a task-specific objective. 

In the seminal paper, Hinton et al.~\cite{hinton2015distilling} proposed to use softened class probabilities from the teacher to provide supervision, which acts as targets for the student model to optimize for. Here, as our teacher model provides $1024$-dimensional embeddings, we instead leverage mean-squared-error loss and a linear layer ($\mathcal{F}_r$) of the same dimensionality on top of student model for distillation purpose~\cite{Romero15-iclr}. We use one-second audio clips as inputs to teacher-student models to get outputs of $1024$ dimensions, on top of which the following loss function is computed:
\begin{equation}
    \mathcal{L}(\theta_{s}, \theta_{r}) = \frac{1}{2} \parallel \mathcal{F}_t(\mathbf{x}; \theta_{t}) - \mathcal{F}_r(\mathcal{F}_s(\mathbf{x}; \theta_{s}); \theta_{r})\parallel^2.
    \label{eq:loss}
\end{equation}
\noindent $\mathbf{x}$ is the training data, $\theta$ represents the respective model parameters, and $\mathcal{F}_r$ is the linear layer model representing a regressor function to match teacher's dimensionality that is discarded after distillation. We use a batch size of $512$ and a fixed learning rate of $0.001$ with an Adam optimizer~\cite{kingma2014adam} to train for approximately $234$K steps with a single NVIDIA RTX3090 GPU.

\subsection{Teacher: Large EfficientNet-V2 Model}
For our teacher model $\mathcal{F}_t$, we use EfficientNet-v2~\cite{tan2021efficientnetv2} from TRILLsson~\cite{shor2022trillsson}, which has achieved exceptional performance on several Non-Semantic Speech Benchmark~\cite{shor20_interspeech} (NOSS) and other related tasks. This model was trained using a combination of two large-scale speech datasets (Speech AudioSet $4.9$K and Libri-light $53$K) via the teacher-student distillation framework. We choose version three of the EfficientNet-v2 model with $21.5$M parameters and access it directly from TensorFlow Hub (TFHub)\footnote{\url{https://tfhub.dev}} during training. This model has a front-end based on log-magnitude Mel spectrograms with $80$ bins ranging from $125$Hz to $7500$Hz, uses window length of $25$ms and hop length of $10$ms, and is initially trained with frame width of $2$s of audio. 

There are larger models available within the TRILLsson family, but we choose the EfficientNet variant as it is both high-performing and less computationally demanding to accommodate a modest compute budget (e.g., hardware  with a single GPU system). This allows us to train, or distill, longer in a short period of time with a large batch size, and demonstrate our central point that we can extract compact binary models purely with knowledge distillation. Further, EfficientNet is a mobile friendly architecture discovered by neural architecture search for image classification tasks with a large capacity. However, we do note that having access to more computing resources may allow one to leverage even bigger models with better supervision, which we leave for future work. 

\subsection{Student: Light-weight Binary Neural Networks}

Our student models $\mathcal{F}_s$ are binary neural network (BNN) architectures with single-bit weights and activations. Because neurons in BNNs can have only two possible states, BNNs provide extreme compression and speed-up gains compared to real-valued artificial neural networks. These fast and energy efficient BNNs are well-suited for deployment on low-resource devices with limited memory and battery power. Specifically, we use binary convolutional networks: a binary DenseNet-28~\cite{bethge2019back} and MeliusNet~\cite{bethge2021meliusnet} with around $4.5$M and $6.4$M parameters, respectively. Their sizes are less than $2$MB in floating-point format, and $1.03$MB for DenseNet-28 and $1.25$MB for MeliusNet in quantized form, which are several folds smaller compared to models utilized in~\cite{shor20_interspeech, shor2022trillsson,peplinski21_interspeech}. We use the Larq~\cite{larq} framework for the implementation of BNNs. 

Our student models are paired with an audio processing front-end based on log-magnitude Mel spectrograms, and can directly consume raw audio waveform. Our front-end uses window size of $25$ms, hop size of $10$ ms, and $64$ Mel-spaced frequency bins in the range of $60$Hz to $7800$Hz for $98$ frames, corresponding to $980$ms. These inputs to the BNNs are then mapped to latent vectors of size $576$ for DenseNet-28 and $512$ for MeliusNet. A GlobalMaxPooling layer on top reduces the size of the bottleneck layer by computing the maximum of all values in each output feature map. Finally, these compact output embeddings are provided to classifiers for downstream tasks. We note that to match TRILLsson's embedding dimensions, we add an additional binary fully-connected layer with $1024$ hidden units $\mathcal{F}_r$, which is discarded after the training phase. 

\section{Experimental Setup}

\subsection{Distillation Dataset}
We perform knowledge distillation with open-source Libri-light dataset~\cite{librilight}, which is derived from public audio books in the LibriVox project. It is the largest publicly available, unlabeled semi-supervised audio dataset to date. From this, we use a medium subset of the dataset with around $5193$ hours of speech (approximately $321$GB in size) due to our modest compute budget. We split each audio clip into non-overlapping one-second segments for training, resulting in around $12$M examples. It is important to note that TRILLsson models are trained with $58$K hours of speech data originating from both Audioset and Libri-light. Also,  TRILLsson is trained with a teacher of massive scale, i.e., CAP12~\cite{shor2022universal}, that is in turn trained on $900$K hours of audio from YT-U data~\cite{zhang2021bigssl}.

\begin{table}[!t]
\centering
\caption{Overview of the downstream evaluation datasets involving non-semantic speech. Number of samples in validation sets are not reported as BRILLsson did not use them.}
\label{tab:data-table}
\resizebox{\columnwidth}{!}{%
\begin{tabular}{@{}lcccc@{}}
\toprule
\multirow{2}{*}{\textbf{Dataset}}   & \multirow{2}{*}{\textbf{Task}}                     & \multicolumn{2}{c}{\textbf{Samples}} & \multirow{2}{*}{\textbf{Classes}} \\ 
& &\textit{Train} & \textit{Test}  &\\
\midrule
MUSAN~\cite{snyder2015musan}            & Speech, Music, and Noise           & 1,613 & 403& 3  \\
ESC-50 (HS)~\cite{piczak2015dataset}  & Human Sounds           & 320 &80  & 10  \\
Voxforge~\cite{maclean2018voxforge}     & Spoken Language   & 117,942 & 30,370 & 6  \\
SpeechCommands~\cite{warden2018speech}  & Commands Recognition  & 85,511 & 4,890  & 12 \\

CREMA-D~\cite{cao2014crema}  & Emotion Detection & 5,144& 1,556  & 6  \\

MSWC-(Micro EN)~\cite{mazumder2021multilingual} & \multirow{2}{*}{Keyword Spotting} & 69,868 & 13,117& 31                                   \\
MSWC-(Micro ES)~\cite{mazumder2021multilingual} &                                    &21,254 & 3,398& 20 \\
Vocalsound~\cite{gong_vocalsound}       & Human Vocal Sounds       &15,570 & 3,594  & 6  \\ \bottomrule
\end{tabular}%
}
\end{table}

\subsection{Downstream Speech Sensing Tasks and Evaluation}
\label{sec:ds_eval}
We evaluate the effectiveness of our method on a broad range of tasks varying from spoken language identification, keyword spotting, accent recognition, identifying emotion, to human vocal sounds monitoring. Table~\ref{tab:data-table} provides an overview and key characteristics of the $8$ datasets. We use MUSAN~\cite{snyder2015musan} to evaluate the detection of music, speech and noise in audio clips. Voxforge~\cite{maclean2018voxforge} is used for identifying spoken English, Spanish, French, German, Russian, and Italian. We use SpeechCommands~\cite{warden2018speech} with $12$ classes for spoken commands and CREMA-D with $6$ classes (anger, disgust, fear, happy/joy, neutral, sad) for emotion recognition. For human sounds task, we utilize $10$ classes subset from ESC-50~\cite{piczak2015dataset}, same as FRILL~\cite{peplinski21_interspeech}. We also use microsets from multilingual word corpus (MSWC) for keyword spotting in English (EN) and Spanish (ES), each task with $31$ and $20$ classes, respectively. Lastly, Vocalsound~\cite{gong_vocalsound} contains audio recording for detection of laughter, sighs, coughs, throat clearing, sneezes, and sniffs.

We follow the published train and test splits of the datasets except for MUSAN, where we randomly split the data into training (80\%) and test (20\%) sets. In case of ESC-50 (HS), following FRILL, we use first four folds as training set and the last fold as a test set. We evaluate the quality of learned representations with a linear classifier trained on top frozen feature extractor or encoder in a similar manner as prior work~\cite{shor20_interspeech, shor2022trillsson}. The classifier is trained with a batch size of $64$ (except for CREMA-D, where we use batch size of $32$ due to relatively small size of the datasets) with learning rate of $0.001$ with Adam optimizer~\cite{kingma2014adam} for $100$ epochs. We use a randomly selected one-second segment from each audio clip in the training set, and evaluate the performance on the entire audio clip during testing. We did not use dev (i.e. validation set) for model selection as the DenseNet and MeliusNet models are already of extremely small size.

\subsection{Latency Benchmarking of Binary Neural Networks}
We use Larq Compute Engine~\cite{larq} for latency benchmarking of our BNN models. To align with prior work, we create a float-$32$ TFLite format model and run it for $150$ runs in a single thread to get an averaged inference time on a device equipped with Snapdragon 855. 

\subsection{Baseline Models for Comparison}
We compare our approach with five methods: TRILL, TRILL-Distilled, FRILL, teacher model TRILLsson3 (EfficientNet-V2), and TRILLsson1 (ResNet-50). TRILL~\cite{shor20_interspeech} is a \textit{TRIpLet-Loss Network} that is pre-trained with large amount of speech data from Audioset. It has shown to learn powerful representations for non-semantic speech tasks and achieved state-of-the-art performance on some of them when it was published in 2020.
It uses a ResNet-50  network architecture, and its layer $19$ has shown to provide the most useful features with dimensionality of $12288$.
TRILL-Distilled~\cite{shor20_interspeech} is a smaller MobileNet-based model with $2048$-dimensional embeddings that is trained with distillation to predict TRILL's embeddings. Along a similar line, FRILL~\cite{peplinski21_interspeech} uses a MobileNetV3 model that is designed to be a fast variant of TRILL specifically focusing on mobile devices. It is trained with distillation to mimic the TRILL layer $19$ representations. In our work, we use \textit{Small 2.0 GAP} model with $2048$-dimensional embeddings due to its best performance. 
Finally, we compare against the teacher model $\mathcal{F}_t$, as well as a smaller model in TRILLsson family, i.e., a ResNet-50  model number one with $5$M parameters. 

We access all the baseline models from TFHub and use them as frozen feature extractors. We use default audio front-end that comes along with the model from TFHub. Similar to our method, we only add a linear classification layer for evaluating performance on downstream tasks as explained in Section~\ref{sec:ds_eval}. In cases where baseline models provide predictions per-time step, we average them to compute final prediction. Furthermore, we train and evaluate baseline models for tasks and datasets that were not presented in the prior work to establish fair comparison; for the rest, we use accuracy score as reported in~\cite{shor2022trillsson,peplinski21_interspeech}. 
The baseline latency values when available are taken from FRILL~\cite{peplinski21_interspeech}. Because the TRILLsson models are released as TFHub modules, we were unable to convert them into the TFLite format and compute on-device latency ourselves in a manner consistent with others. Importantly, unlike previous works that trained multiple linear classifiers with different techniques, we only train and evaluate a logistic regressor implemented with a linear dense layer. An exhaustive search over classification methods may yield further improvement.

\section{Results and Discussion}
\label{sec:results}
We evaluate the performance of our BRILLsson approach, and contextualize how well binary models generalize on a broad range of speech sensing tasks as compared to large-scale models. Table~\ref{tab:main-table} presents results of BRILLsson along with five large-scale baselines models. Since most of the datasets are published with fixed train-test-validation splits, the convention for audio recognition is to use the validation set for model selection and report a single classifier's performance on the test set. Since these models are trained and tested on the same samples, the accuracies in Table 2 can be directly compared to each other without error bars. First, we notice that the teacher model TRILLsson (EN-v2) overall performs well in comparison to other floating-point based models, which highlights the usefulness of this model as teacher for distillation.

\renewcommand{\arraystretch}{1.5}
\begin{table*}[!t]
\centering
\caption{Generalization performance of BRILLsson on a range of non-semantic speech representations tasks. We train a linear classifier on top of distilled frozen models to assess the quality of learned embeddings. In all cases embeddings are from the last layer of the corresponding BRILLsson models.
RN-50 is ResNet-50, EN-v2 is EfficientNet-v2, DN is DenseNet-28, and MN is MeliusNet. T denotes a tiny model based on a intermediate layer of DenseNet-28. 
}
\label{tab:main-table}
\resizebox{\textwidth}{!}{%
\begin{tabular}{@{}lcccccccccc@{}}
\toprule
\textbf{Method} &
  \multicolumn{1}{l}{\textbf{MUSAN}} &
  \multicolumn{1}{l}{\textbf{ESC-50 (HS)}} &
  \textbf{Voxforge} &
  \textbf{SpeechCommands} &
  \textbf{CREMA-D} &
  \textbf{MSWC-EN} &
  \textbf{MSWC-ES} &
  \multicolumn{1}{l}{\textbf{Vocalsound}} &
  \textbf{Size (MB)} &
  \textbf{Latency (ms)} \\ \midrule
TRILL             & 98.2 & 86.4 & 84.5 & 81.9 & 66.2 & 81.3 & 88.0 & 88.2 & 98.1  & 275.3              \\
TRILL-Distilled   & 98.5 & 87.9 & 80.0 & 80.2 & 70.2 & 74.4 & 87.9 & 85.8 & 107.1 & 22.5               \\
FRILL             & 98.2 & 86.4 & 76.9 & 79.7 & 70.9 & 79.1 & 87.6 & 86.7 & 38.5  & 8.5                \\
TRILLsson (RN-50) & 98.5 & 60.0 & 98.6 & 91.2 & 81.3 & 91.4 & 94.5 & 87.2 & 22.0  & \multirow{2}{*}{-} \\
TRILLsson (EN-v2) & 98.7 & 87.5 & 99.2 & 93.2 & 83.2 & 87.2 & 93.9 & 89.0 & 99.0  &                    \\ \midrule
BRILLsson (DN)    & 93.0 & 85.0 & 70.8 & 88.7 & 65.3 & 87.6 & 88.6 & 80.2 & 2.0   & 6.4                \\
BRILLsson (MN)    & 91.5 & 80.0 & 70.1 & 89.2 & 63.8 & 88.5 & 89.1 & 83.2 & 2.1   & 7.6                \\
BRILLsson (T)     & 90.5 & 73.7 & 73.0 & 88.5 & 54.6 & 87.2 & 88.4 & 78.4 & 0.65  & 6.1                \\ \bottomrule
\end{tabular}%
}

\end{table*}

Our DenseNet (DN) based binary model demonstrate excellent performance on all considered tasks even with its small size. Note that DN is only $2$MB including the audio front-end, whereas the teacher model has size of around $99$MB. Similarly, our MeliusNet (MN) has similar or slightly better performance than DN, in particular on Vocalsound where it achieves accuracy of $83.2$\%. This is only 2\% less than the acuracy of TRILL-Distilled. Interestingly, BRILLsson has superior generalization on keyword spotting tasks, achieving $89.2$\% accuracy on SpeechCommands and $88.5$\% on MSWC-EN. However, BRILLsson saw performance downgrade in CREMA-D and Voxforge. For CREMA-D, we believe the limited capacity of our model, challenging nature of the problem (i.e. emotion detection) and small dataset are the main reasons that BRILLsson does not perform as well. We hypothesize that scaling the dataset may improve the performance of classifier trained on top of BRILLsson.
On the other hand, Voxforge has more samples per class than CREMA-D, but spoken language identification is still a very difficult task. In particular, the pre-training dataset we used contains only English speech, whereas Voxforge contains other languages like German. We believe the main difficulty lies in the network trying to extrapolate to these audio inputs that it did not see during pre-training. We once again emphasize that our BRILLsson models have latency of less than $8$ms with size of merely around $2$MB.

Furthermore, we add a linear classifier after \textit{batch-normalization-12} intermediate layer in DN model to experiment with creating an even smaller model, labeled BRILLsson (T) in Table~\ref{tab:main-table}. The resulting model has size of $0.65$MB and latency of $6.1$ms. Interestingly, on Voxforge the tiny model achieves $73.0$\% while having only $1.4$M parameters.   These results demonstrate the usefulness of representations learned with BNNs and that a single BNN model can be used as a feature extractor on low-resource devices for multiple downstream tasks.  

\begin{figure}[ht]
    \centering
    \includegraphics[width=0.6\columnwidth]{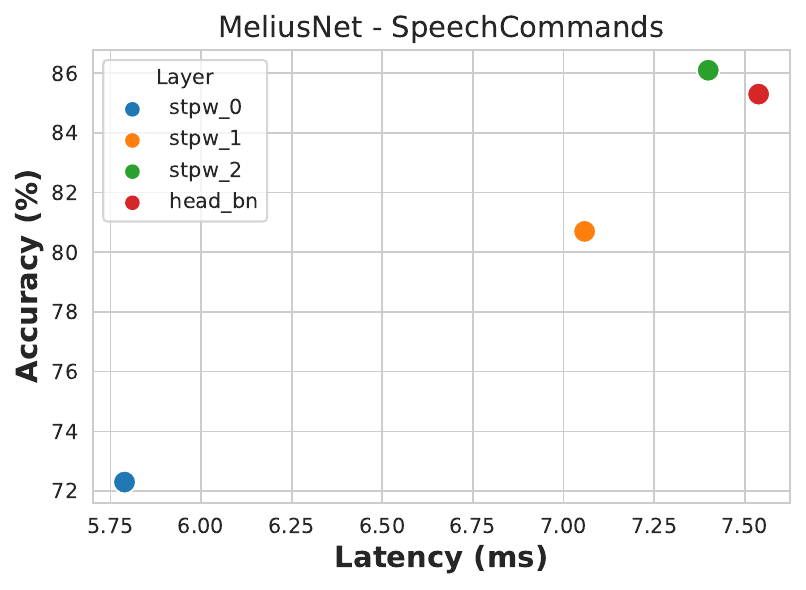}
    \caption{Performance of intermediate layers' representations on SpeechCommands for MeliusNet TFLite format model. Each point corresponds to the accuaracy and latency of a model where we added a classifier after an intermediate layer. For example, a model with classifier after \textit{section-1-transition-pw} layer achieves $81$\% accuracy with latency of $7$ms. \textit{stpw} is an abbreviation of \textit{section-transition-pw}. \textit{head\_bn} is the final layer of MN.}
    \label{fig:meliusnet_spc}
\end{figure}

Along similar lines, we explore the quality of representations from intermediate layers of the distilled model using MN on SpeechCommands. For each intermediate layer, a classification head is added on top and trained in the same manner as previous experiments. The rest of the model is fixed during this phase. Then, we convert each model to TFLite format, evaluate its accuracy and latency, and report the results in Figure~\ref{fig:meliusnet_spc}. We see that conversion to TFLite format does not result in a significant performance degradation. Also, we observe a trade-off between accuracy and speed, as expected; for instance, the model built on the layer \textit{section-2-transition-pw} has the highest accuracy of $86.1$\%, but has relatively high latency of $7.3$ms. A more comprehensive investigation into the optimal BNN architecture and intermediate layer embeddings is deferred to future work.

\section{Conclusions}
We have designed, developed, and publicly released  BRILLsson: an extremely compact, fast, and flexible model for non-semantic speech representation learning. We used embedding distillation to transfer knowledge from an existing TRILLsson model to small binary neural networks. Our approach significantly reduced the model size while keeping the performance on par with large-scale real-valued counterparts, which is valuable for low-resource devices. While this work focused on utilizing existing neural architectures, we would like to explore neural architecture search methods in the future to design even more light-weight BNN models that are suitable for micro controllers.

\bibliographystyle{plain}
\bibliography{ms}

\end{document}